\documentstyle [12pt,fleqn]{article}
\begin{document}
\begin{titlepage}
\vspace*{-15 pt}
\begin{flushright}
EFI 90-62 \\
August 1991\\
\end{flushright}
\begin{center}
\vspace{.25 in}
{\large \bf The renormalization group flow in 2D  N=2 SUSY Landau-Ginsburg
 models\footnote{Work supported in part by DOE grant DE-FG02-90ER40560.}}

\vspace{.3 in}
\centerline{Jadwiga Bie\'{n}kowska}
\vspace{.6 cm}
{\em Department of Physics and \\
 Enrico Fermi Institute\\
 The
University of Chicago \\

   Chicago, IL 60637 }
\end{center}
\vspace{.6 cm}
\baselineskip=18pt
\centerline{\bf Abstract}
\vspace{-1 cm}
\begin{quote}
\noindent

We investigate the renormalization of N=2 SUSY
L-G models with  central
charge $c=3p/(2+p)$ perturbed by an almost marginal chiral
operator.  We calculate the
renormalization of the chiral fields up to $gg{^*}$ order
and of nonchiral fields up to $g(g^{*})$ order. We propose a formulation of
the nonrenormalization theorem and show that it holds in the lowest
nontrivial order. It turns out that, in this approximation, the chiral fields
can not get renormalized $\Phi^{k}=\Phi^{k}_{0}$. The  $\beta$ function  then
remains unchanged $\beta=\epsilon g$.
\end{quote}\end{titlepage}
 \vspace{1.5 in}
\pagebreak
\section{Introduction}
A better understanding of the structure of N=2 SUSY
theories in two dimensions is important from two points
of view. First the N=2 SUSY conformal theories can be
regarded as  building blocks for the background
``spacetime'' in  string theories with N=1 SUSY in  four dimensions
\cite{rg:fl1} , \cite{rg:fl2}.
 Second they are of great interest from the point of view of statistical
systems where N=2 conformal field theories can be understood as two-dimensional
self-dual critical points in $Z_{N}$-symmetric
statistical systems [4]. The content of the minimal model series
with central charge $c=3p/(2+p)$ $p=2,3 \ldots$,
i.e. the classification of primary fields and fusion
rules, was extensively derived by several authors
\cite{rg:fl4},\cite{rg:fl5},\cite{rg:fl6}. There are also interesting
connections between the N=2 minimal models and singularity theory which have
been dicovered recently \cite{rg:fl12}, \cite{rg:fl13}. The
 structure beyond the conformally invariant point of
the theory still remains  open for investigation.

In the space of all two dimensional N=2 SUSY theories,  minimal models
are  fixed points of the renormalization group flow with $c<3$ \cite{rg:fl2},
\cite{rg:fl3}, \cite{rg:fl4}. In our paper we try to describe the structure of
this space by investigating the RG-flow near  the fixed
point. Different   N=2 minimal models
 have a different value of the supersymmetry index, so passing from one
theory to the other can not occur by a smooth and finite change
of parameters as it can  for N=0,1 SUSY  \cite{rg:fl3}, \cite{rg:fl7},
\cite{rg:fl8}, \cite{rg:fl9}. It was pointed out by Cveti\v{c}
and Kutasov that the fixed points of N=2 SUSY are infinitely apart from each
other in  the space of coupling constants where the metric tensor is defined by
the Zamolodchikov \cite{rg:fl8} formula $G_{ij}(g)= \langle
\Phi_{i}(g)\Phi_{j}(g)\rangle_{g}$. We confirm this result  by a perturbative
calculation.

Another interesting issue   is the nonrenormalization
conjecture \cite{rg:fl10}, \cite{rg:fl11} - the absence on nontrivial
renormalization of the superpotential. Ordinarily this conjecture is proven in
the loop expansion, and it is not clear whether it survives the infrared
regularization; renormalization and resummations necessary to define the
nontrivial Landau-Ginsburg fixed point.
 Existing proofs of the nonrenormalization
theorem \cite{rg:fl10}, \cite{rg:fl11} are based on the structure of the
graphical expansion about free field theory. This expansion is badly infrared
divergent in a massless Landau-Ginsburg model; these divergences must be
resummed and the effect on the nonrenormalization theorem is unclear. One can
expect that nonrenormalization should enforce the existence of only one
renormalization constant (the wave function renormalization) for all fields in
the theory. Stating it differently we could expect that renormalized fields
$\tilde{\Phi^k}$ would be again powers of some basic field $\tilde{\Phi}$.

Indeed it is not entirely clear what
the precise statement of nonrenormalization is. For instance there always
exist local coordinates near a fixed point in which the $\beta$ function is
linear \cite{rg:fl17}, moreover any statement about the form of the
$\beta$ function can be changed by redefinition of couplings. Such coordinates
are usually singular (they are essentially the bare coordinates) and
put all nontrivial fixed points at coordinate infinity ( although they might be
a finite physical distance in the Zamolodchikov metric; in any case for our
situation the result of \cite{rg:fl7} indicates that there are no nearby fixed
points). One might say that nonrenormalization is the statement  that
there are no divergent counterterms needed to renormalise the chiral fields
sector, so by a {\em finite} renormalization (nonsingular coordinate
transformation) one can bring the chiral fields $\beta$ function into a
linearized form. We  will calculate  this $\beta $ function in the composite
operator (conformal) perturbation theory scheme of Zamolodchikov, and show
that it is the scheme for which this form of nonrenormalization holds to
leading nontrivial order.

In this paper we look closely at N=2 SUSY  minimal
theories described in the Landau-Ginsburg formulation \cite{rg:fl2},
\cite{rg:fl3} by one chiral field $\Phi$ where the lagrangian is

\begin{equation}
S_{0}=\int d^{2}zd^{2}\theta d^{2}\theta^{*}\Phi\Phi^{*}+\int
d^{2}zd^{2}\theta \Phi^{p+2} + \int
d^{2}zd^{2}\theta^{*}\Phi^{*(p+2)} \label{1}
\end{equation}
at the conformally invariant point with   the central charge value
$c=3p/(2+p)$. We perturb the L-G model (~\ref{1}) by the most marginal operator
$\Phi^{p}$. Solving the perturbed theory up to $gg^{*}$ order for
chiral fields we prove, in leading nontrivial order, that the
nonrenormalization
theorem holds for chiral fields and their coupling
constants in a trivial way $\tilde{\Phi}^{k}=\Phi^{k}$.  The
renormalization of nonchiral fields does not obey a
simple relation.

We organize our paper as follows. In the first
section we recall the LG model together with its
field content. In section 3 we describe the geometrical
approach to the RG-flow and point out features specific
to a chiral theory. In section 4 we present
perturbative results for nonchiral fields and in section 5 we discuss the
renormalization of chiral fields  and couplings and calculate the $\beta$
function in this approximation.

While the present work was being completed, we received a preprint by
W.A.Leaf-Hermann \cite{rg:fl25} investigating substantially the same problem,
we comment on his results in the discussion.

 \section{N=2 SUSY
Landau-Ginsburg models}
The conformally invariant point of N=2 SUSY discrete
series with  the central charge value $c=3p/(2+p)$
 $p=2,3\ldots$ can be regarded as a L-G model depending on the
single chiral field $\Phi$ and its hermitian conjugate
$\Phi^{*}$ \cite{rg:fl2}, \cite{rg:fl3}. A general
complex SUSY field is a function on the N=2 superspace with
local coordinates
$Z=(z,\theta,\theta^{*})$,
$\bar{Z}=(\bar{z},\bar{\theta},\bar{\theta}^{*})$. The
charge of super variables $\theta$, $\theta^{*}$ is
$q=\pm1/2$. The covariant derivatives  in these
coordinates are
\begin{eqnarray}
D=\frac{\partial}{\partial\theta^{*}}
 +\theta \frac{\partial}{\partial z}\nonumber\\
D^{*}=\frac{\partial}{\partial\theta}+
 \theta^{*}\frac{\partial}{\partial z} \label{2}
\end{eqnarray}
The N=2 (anti)chiral fields $\Phi$, $\Phi^{*}$ are defined by the conditions
\begin{eqnarray}
D^{*}\Phi^{*}=\overline{D^{*}}\Phi^{*}=0 \nonumber\\
D\Phi=\overline{D}\Phi=0 \label{3}
\end{eqnarray}
The conformally invariant L-G model is described by the action (~\ref{1}),
where
for the dependence of the chiral field on the superspace
parameters we  use the shorthand notation
$\Phi(z,\theta,\bar{z},\bar{\theta})=\Phi(Z,\bar{Z})$ and
respectively
$\Phi^{*}(z,\theta^{*},\bar{z},\bar{\theta}^{*})=\Phi^{*}(Z^{*},\bar{Z}^{*})$.
 Following the Zamolodchikov prescription \cite{rg:fl2}, \cite{rg:fl3},
\cite{rg:fl14} we
identify different composite fields of $\Phi$, $\Phi^{*}$ with the
primary fields of N=2 SUSY theories already classified \cite{rg:fl4},
\cite{rg:fl5},
 \cite{rg:fl6} . For our calculations we find useful the identification of N=2
with parafermion models (PF) \cite{rg:fl4}, \cite{rg:fl6}. In terms of
 parafermion fields $\phi$ and  free field $\varphi$
lowest components of the holomorphic part of (anti)chiral fields
 ($\Phi^{*k}$),$\Phi^{k}$ look like
\begin{eqnarray}
\Phi^{k}(z) \approx \phi^{k}_{k}(z):exp\left( i\frac{k\varphi
(z)}{\sqrt{2p(p+2)}}\right):\nonumber \\
\Phi^{*k}(z) \approx  \phi^{k}_{-k}(z):exp\left(i\frac{-k\varphi
(z)}{\sqrt{2p(p+2)}}\right):
 \label(4)
 \end{eqnarray}
where the conformal dimension and charge are given by
\begin{eqnarray}
\Delta &=& q=\frac{k}{2(p+2)} \nonumber \\
\Delta &=& -q=\frac{k}{2(p+2)}\,\,\,\,\,\, k=1,\ldots ,p \label{5}
\end{eqnarray}
The antiholomorphic part dependent on $\bar{z}$ has the same form as above.
 As a consequence of the
equation of motion $D\bar{D}\Phi^{*}=(p+2)\Phi^{p+1}$, the (p+1) power of the
chiral field $\Phi$ is not  primary.
 Various composites of the chiral-antichiral
fields are expressed in terms of parafermions by the identification of the
lowest component
\begin{equation}
\Phi^{k}\Phi^{*l}(z)\approx
\phi^{j}_{m}(z):exp\left(i\frac{m\varphi (z)}{\sqrt{2p(p+2)}}\right): \label{6}
\end{equation}
where $m=k-l$ and $j=min(k+m,2p-(k+m))$ and  the conformal dimension and
charge are given by
\begin{equation}
\Delta=\frac{j(j+2)-m^{2}}{4(p+2)}, \,\,\, q=\frac{m}{2(p+2)} \label{7}
\end{equation}
 The highest components of the field are defined by the operator product
expansion of  lowest components with the supersymmetry generators which are
expressed in terms of parafermionic currents $\psi^{\pm}_{1}$ by
\begin{equation}
G^{\pm}_{1/2}(z)=\sqrt{\frac{2p}{p+2}}\psi^{\pm}_{1}(z):exp\left(\pm
i\frac{(p+2)\varphi (z)}{\sqrt{2p(p+2)}}\right): \label{8}
 \end{equation}
To complete the set of N=2 SUSY generators we recall that the U(1) current is
given by \begin{equation}
J(z)=\frac{i}{2}\frac{p}{\sqrt{2p(2+p)}}\partial \varphi (z) \label{9}
\end{equation}
and the stress energy tensor is
\begin{equation}
T(z)= T_{PF}-\frac{1}{4} :\partial \varphi ^{2}(z): \label{10}
\end{equation}
The fusion rules of the N=2 SUSY and the PF model are in ref. \cite{rg:fl4},
 \cite{rg:fl6}, as well as
 N=2 SUSY Ward identities and the general form of the three point
function that we use in the perturbative calculation.  The normalization
of the fields is set by the normalization of the two point function which
we choose to be
\begin{equation}
 \langle \Phi^{l_{1}}_{m_{1}}(Z_{1})\Phi^{*l_{2}}_{m_{2}}(Z_{2})\rangle=
 Z_{12}^{-2\Delta_{1}}
(1+2q_{1}\frac{\theta^{*}_{12}\theta_{12}}{Z_{12}})
\delta_{l_{1},l_{2}}\delta_{m_{1},m_{2}} \label{11}
\end{equation}
  for nonchiral fields and
\begin{equation}
\langle\Phi^{l_{1}}_{l_{1}}(Z_{1})\Phi^{*l_{2}}_{-l_{2}}(Z^{*}_{2})\rangle=
Z_{12}^{-2\Delta_{1}}
(1+2q_{1}\frac{\theta_{1}\theta^{*}_{2}}{Z_{12}})
\delta_{l_{1},l_{2}} \label{12}
\end{equation}
for chiral fields, where $Z_{12}= z_{12} -\theta_{1}\theta^{*}_{2}
-\theta_{2}\theta^{*}_{1}$ for nonchiral two point function and $Z_{12}= z_{12}
-\theta_{1}\theta^{*}_{2}$ for chiral fields.
 In this model the relevant perturbations  are
those by chiral fields $\Phi^{k}$ \cite{rg:fl3}. All nonchiral fields are
irrelevant since their dimensions are always greater than zero and so the
dimensions of the highest component are always $\Delta>1$. This is also the
reason why we can treat the coupling constant space as finite dimensional.

 To have some grasp on the RG-flow in these theories we investigate
the model (~\ref{1}) perturbed by the almost marginal (for large $p$ ) operator
$\Phi^{p}$. The complete action we consider  is
\begin{equation}
S=\frac{1}{2 \pi}\left( S_{0}
+g\int  d^{2}zd^{2}\theta\Phi^{p}+g^{*}\int d^{2}zd^{2}\theta^{*}\Phi^{*p}
\right)
 \label {13} \end{equation}
where $1/2\pi$ is the normalization of the action needed to define correctly
the
stress-energy tensor (see \cite{rg:fl3}, \cite{rg:fl8},
\cite{rg:fl9}, \cite{rg:fl16}). The dimension of the
$\Phi^{p}$ field is $\Delta_{p}=1/2-\epsilon$, where $\epsilon=1/(p+2)$. Since
integrated fields are charged the coupling constant charge (left plus right)
can be identified as $q(g)=2/(p+2)=2\epsilon$.

\section{Geometrical description of RG-flow in N=2 SUSY theories}
In this section we recall the notation of the geometrical approach to
the RG-flow  developed in recent years  \cite{rg:fl8},
\cite{rg:fl16}. We point out
some facts which are specific to  N=2 SUSY chiral theories.

The field theory can be considered as a set of (connected) correlation
functions
\begin{equation}
\langle\Phi_{1}(z_{1})\ldots \Phi_{N}(z_{N})\rangle\,, \label{14}
\end{equation}
where the local fields are elements of the infinite dimensional vector space
and
form a closed operator algebra.  N-point correlation functions can be
regarded as maps $\otimes_{n}\cal A \rightarrow \bf R$, i.e.
tensors of rank $(0,N)$. The correlation functions depend on a
number of   coupling constants $g^{i}$. Different
descriptions of a system, corresponding for example to different choices of
renormalization scales,
are related to each other by a continuous change of $g$'s. The $g^{i}$ can be
regarded as  coordinates on the state manifold $\cal M$ of the given system
and  equivalent coordinate systems are related by diffeomorphisms.

The correlation function (~\ref{14}) can be derived from the Lagrangian
formulation of the theory by the functional integral
\begin{equation}
\int{\cal D} \Phi \,\,\Phi_{1}(z_{1})\ldots \Phi_{N}(z_{N})\, e^{-S(\Phi)}
\label{15}
\end{equation}
The decomposition $S=S_{0}+S_{prt}$, as in (~\ref{13}), corresponds to the
choice of bare coordinates $g^{i}_{0}$ in which $S_{0}$ is the unperturbed
action at the point $g^{i}_{0}=0$ and $S_{prt}=g^{i}_{0}\int
d^{2}Z\Phi_{0i}+h.c.$ . In this formulation the space of integrated fields
$\Phi_{0i}=\int d^{2}Z \Phi_{0i}(Z,\bar{Z})$ can be identified with the  space
tangent
to the manifold $\cal M$ \cite{rg:fl8},
\cite{rg:fl16}. For example, in the bare coordinate system the
derivatives $\frac{\partial}{\partial g^{i}_{0}}$ acting on the correlation
function are
\begin{equation}
\frac{\partial}{\partial g^{i}_{0}}\langle\Phi_{01}(z_{1})\ldots
\Phi_{0N}(z_{N})\rangle =
\langle \Phi_{0i}\,\Phi_{01}(z_{1})\ldots \Phi_{0N}(z_{N})\rangle\,.
\label{16}
\end{equation}

The RG-flow is a one-parameter group of the action space diffeomorphisms with
group parameter $t=\ln(\frac{a}{R})$,where $a$ is the scale of the system(i.e.
lattice spacing) and $R$ is the scale of the observer
\cite{rg:fl16}. The generating
vector field is $\Theta =\beta ^{i}\Phi_{i}$, the integrated trace of
stress-energy tensor, and the components $\beta ^{i}=\frac{d}{dt} g^{i}$ are
called beta functions; $\beta ^{i}$ depend on $g$ - the choice of coordinate
system. In the neighborhood of a fixed point $g^{i*}$ one can introduce the
bare coordinates which linearize the  beta functions \cite{rg:fl17}
\begin{equation} \beta_{0}^{i}=g_{0}^{j}\gamma ^{*i}_{j}\,,
\label{17}
\end {equation}
where $\gamma^{*i}_{j}$ is the matrix of anomalous dimensions at the fixed
point. The RG-flow is described in  generic  coordinates $g$ by
\begin{eqnarray}
g^{i}=Z^{i}_{j}(g)g^{j}_{0} \nonumber \\
\Phi_{i}=\Phi_{0j}\tilde{Z} ^{j}_{i}(g)\,,
\label{18}
\end{eqnarray}
where $g^{i}$ and $\Phi_{i}$ are renormalized couplings and fields. The
$Z^{i}_{j}(g),\tilde{Z} ^{j}_{i}(g)$ functions determine beta functions and the
anomalous dimension
matrix \cite{rg:fl16}.

The requirement that the RG-flow  leaves the theory invariant imposes
constraints on  correlation functions which can be written down as the
Callan-Szymanzik equation \cite{rg:fl8}, \cite{rg:fl9}
 \begin{equation}
\left[\sum_{a=1}^{N}\frac{1}{2} Z^ {a}_{i}\frac{\partial}{\partial
Z^{a}_{i}} +\hat{\Gamma}(g) -\sum_{i=1}^{p}\beta
^{j}(g)\frac{\partial}{\partial g^{j}}\right] \langle \Phi_{1}(z_{1})\ldots
\Phi_{N}(z_{N})\rangle=0\,.
\label{19}
 \end{equation}
$\hat{\Gamma}(g)$ is the matrix of anomalous dimensions acting on fields as
$\hat{\Gamma}(g)\Phi_{i}=\gamma ^{j}_{i} \Phi _{j}$ and
\begin{equation}
Z^ {a}_{i}\frac{\partial}{\partial Z^{a}_{i}}=z^
{a}\frac{\partial}{\partial z^{a}}+ \frac{1}{2}\theta ^
{a}\frac{\partial}{\partial \theta ^{a}} +c.c.\nonumber
\label{20}
 \end{equation}
for chiral fields and respectively $Z^{*}$ for antichiral fields and
\begin{equation}
Z^ {a}_{i}\frac{\partial}{\partial Z^{a}_{i}}=z^{a}
\frac{\partial}{\partial
z^{a}}+ \frac{1}{2}\theta ^ {*a}\frac{\partial}{\partial \theta
^{*a}}+\frac{1}{2}  \theta ^ {a} \frac{\partial}{\partial
\theta ^{a}}+c.c.
\label{21}
\end{equation}
 for nonchiral fields.

 The most general  relevant perturbation of (~\ref{1}) is
\begin{equation}
S_{prt}=g^{k}\int d^{2}Z\Phi_{k} + h.c.
\label{22}
\end{equation}
with   coupling constants having charges
$q(g^{k})=-q(g^{*k})=\frac{(p+2-k)}{(p+2)}=2\epsilon _{k}$ where
$\epsilon_{k}\ll1$ for $1\ll k$. At $g^{i}=0$ the  space tangent to the
manifold of chiral coupling constants has a basis of integrated chiral fields
$\Phi_{i}=\int d^{2}Z\Phi_{i}(Z,\bar{Z})$ with the hermitian scalar product
\begin{equation}
(\Phi_{i}|\Phi^{*}_{j})=\delta _{i\bar{j}}\,.
\label{23}
\end{equation}
 The metric   on the manifold $\cal M$ is $G_{i\bar{j}}
dg^{*i}dg^{\bar{j}}=d^{2}s$ . As a consequence of the hermitian structure, the
renormalization
transformations (~\ref{18}) should be also hermitian.

Looking at the N=2 SUSY   fusion
rules \cite{rg:fl4},
\cite{rg:fl6}, we can find that in general fields renormalize as follows
\begin{eqnarray}
\tilde{\Phi}^{k} &=& \Phi ^{k} + O(gg^{*})\Phi^{k} +O(g^{3}) \label{24} \\
\tilde{\Phi}^{l}_{m} &=& \Phi ^{l}_{m} + O(g)\Phi^{l}_{m-2} +
O(g^{*})\Phi^{l}_{m+2}+ O(g^{2})
\label{25}
\end{eqnarray}

 The condition (~\ref{24}) is justified as the correct renormalization
procedure for this  perturbation if one considers the three point functions of
chiral fields which includes $\Phi^{p}$ field. From the N=2 SUSY Ward
identities one can easily find that the only nonvanishing three point function
could be
  $\langle\Phi^{k}\Phi^{*(k-2)}\Phi^{p}\rangle= CF(\theta_{i},z_{i})$ [6].
The fact that the lowest component of the $\Phi^{p}$ is free scalar field
allows us to calculate the $C$ constant easily and one can check that it is
$C\approx  \langle\phi^{k}_{k}\phi^{-(k-2)}_{k}\rangle=0$. Therefore
all  three point functions involving the $\Phi^{p}$ field vanish. This
leads us to set the normalization condition for this perturbation theory as in
equation (~\ref{24}) if we choose the coordinate system with $\partial_{k}
G_{ij}=0$. Than the metric for chiral fields will have
 the form
\begin{equation} G_{i\bar{j}}=\delta _{i\bar{j}} + O(gg^{*}) \delta _{i\bar{j}}
\label{25c}
\end{equation}
 Since the second derivatives of the metric (at $g=0$) determine the
curvature tensor in the frame in which   $\partial_{k}
G_{ij}=0$, we  have to look closely at the renormalization presciption for
the chiral field in the next paragraphs.
 For nonchiral fields we use the condition of
renormalization set down by the requirement that the renormalized metric is
\begin{equation} G_{ij}=\delta _{ij} + O(gg^{*}).
 \label{27}
\end{equation}

\section{Renormalization of nonchiral fields}
It follows from the N=2 SUSY Ward identities \cite{rg:fl6}  that the only
nonvanishing three point functions of $\Phi_{1},\Phi_{2},\Phi_{3}$  are the
ones for which either $q_{1}+q_{2}+q_{3}=0$ or $q_{1}+q_{2}+q_{3}=\pm
\frac{1}{2}$ . This fact coincides with the soft breaking of the U(1)-chiral
current symmetry \cite{rg:fl15}. According to these properties
renormalized fields should have  exactly the same  charge as the unrenormalized
fields, when expressed in terms of fixed point operators and the charge of the
couplings is accounted
for.

Using these two facts we get the general formula (in the perturbed theory
(~\ref{13})) for the renormalization of nonchiral fields in the first order in
$g(g^{*})$ \begin{equation}
\tilde{\Phi} ^{l}_{m}=\Phi ^{l}_{m}+g A^{l}_{m,m-2}\Phi ^{l}_{m-2}+
g^{*} A^{l}_{m,m+2}\Phi ^{l}_{m+2} + O(g^{2})\,.
\label{28}
\end{equation}
All fields in  the above formula have to be nonchiral.
For notational convenience we set
\begin{equation}
\Phi^{p}=\Phi_{1}\, \, , \,
\Phi^{l}_{m}=\Phi_{2}\,\, , \,
\Phi^{l}_{m-2}=\Phi_{3}\,.
\label{29}
\end{equation}
 According to  fusion rules the
chiral fields $\Phi^{l}_{l}$ could  renormalize by  mixing with
$D\bar{D}\Phi^{l}_{l\pm2}$, but these fields have dimensions differing by
a number close to $\frac{1}{2}$ and so such mixing is not possible
\cite{rg:fl8}. Fields $\Phi^{1}_{\pm1}$ do not mix either, as it is  easly
seen  from the parafermion fusion rules and the N=2 SUSY Ward identities
\cite{rg:fl4}, \cite{rg:fl6}.
The requirement that the metric has the form (~\ref{27}) is equivalent to the
condition
\begin{equation}
\frac{\partial}{\partial g}\langle\Phi_{2}(1)\Phi_{3}(0)\rangle _{|g=0}=0
\label{30}
\end{equation}
It is enough to consider the above equation only for two nonchiral fields,
because fields
$\Phi^{l}_{m+2}$, $\Phi^{l}_{m-2}$ do not mix with each other. Following
perturbation theory methods we get that
\begin{eqnarray}
A_{23}=\frac{1}{2}C(\Gamma ^{2}_{13}+\Gamma ^{1}_{23}-\Gamma ^{3}_{12})
\nonumber \\
A_{32}=\frac{1}{2}C(\Gamma ^{3}_{12}+\Gamma ^{1}_{23}-\Gamma ^{2}_{13})
\label{31}
\end{eqnarray}
where
\begin{eqnarray}
\Gamma ^{1}_{23}&=&\frac{1}{2\pi}|z_{2}-z_{3}|^{2(\Delta_{2}+\Delta_{3}+2
-\epsilon)}\int d^{4}Z_{1} d^{4}\theta _{2}
d^{4}\theta _{3}
\langle \Phi_{1}(Z_{1})\Phi_{2}(Z_{2})\Phi_{3}(Z_{3})\rangle \nonumber \\
&=&-\frac{\Gamma(1-2\epsilon)\Gamma (\epsilon+\Delta_{3}-\Delta_{2})
\Gamma (\epsilon+\Delta_{2}-\Delta_{3})}
 {2\Gamma(2\epsilon)\Gamma(1-\epsilon+\Delta_{3}-\Delta_{2})
\Gamma(1-\epsilon+\Delta_{2}-\Delta_{3})}\times \nonumber\\
& \times
&[(1+\Delta_{2}+\Delta_{3}-\epsilon)(\Delta_{2}+\Delta_{3}-\epsilon)]^{2}
\nonumber \\  \Gamma ^{2}_{13}&=&-2\frac{\Gamma
(\epsilon-\Delta_{2}+\Delta_{3})
\Gamma(1-\epsilon-\Delta_{3}-\Delta_{2})\Gamma(2\Delta_{2})}
{\Gamma(1-\epsilon+\Delta_{2}-\Delta_{3})
\Gamma(\epsilon+\Delta_{3}+\Delta_{2})\Gamma(1-2\Delta_{2})}\times
\nonumber \\ & \times &
[(1+\Delta_{2}+\Delta_{3}-\epsilon)(\Delta_{2}-q_{2})]^{2} \,.
\label{32}
\end{eqnarray}

$\Gamma ^{2}_{13}$ is obtained in a similar way as $\Gamma ^{1}_{23}$
by integration over all variables corresponding to the field
$\Phi_{2}=\Phi^{l}_{m}$ and $\Gamma ^{3}_{12}$ is obtained by the exchange of
indices $2\leftrightarrow 3$ in  $\Gamma^{2}_{13}$. $\Delta$, $q$ are
respectively dimensions and charges of
 fields. C is a constant derived from fusion rules and  the normalization
of fields (~\ref{11}), (~\ref{12}) $C=\frac{p}{p+2}
(2\Delta_{2}\Delta_{3}(2\Delta_{2}+1)(2\Delta_{3}+1))^{-1}$ (in our definition
of $\Gamma's$ we use the general form of three point function derived directly
from N=2 SUSY Ward identities [6] which is defined up to a multiplicative
constant) . Under the perturbation (~\ref{13}) the space of fields
$\Phi^{l}_{m}$ (for each $l$) is divided into two invariant subspaces of $m$
odd or even. Within each  subspace one can write the formula for the anomalous
dimension matrix \cite{rg:fl8} (fields $\Phi^{l}_{l}, \Phi^{l}_{-l}$, being
chiral, are excluded) \begin{equation}
 \gamma^{i}_{j}(g)=\left(
\begin{array}{ccccc}
\ddots & . & 0 &  0 & 0 \\
. & \Delta_{m+2} & g\gamma_{m+2,\overline{m}} & 0 & 0 \\
0 & g^{*}\gamma_{m,\overline{m+2}} & \Delta_{m} & g\gamma_{m,\overline{m-2}} &
0
 \\
 0 & 0 & g^{*}\gamma_{m-2,\overline{m}} & \Delta_{m-2} & .\\
0 & 0 & 0 & . & \ddots
\end{array}
\right)
\label{33}
\end{equation}

where $\gamma_{m,\overline{m-2}}=\gamma_{m-2,\overline{m}}=\frac{1}{2}C
[\epsilon \Gamma^{1}_{23} +(\Delta_{3} -\Delta_{2})(\Gamma^{2}_{13}
-\Gamma^{3}_{12})]$.

In the general case it is not easy to find the eigenvalues of
this matrix, but one can see that it is hermitian and that the eigenvalues will
be of order $gg^{*}$,  $\tilde{\Delta} _{m}= \Delta_{m} +O(gg^{*})$, as  could
be expected.

\section{Renormalization of chiral fields}
Considering the U(1) charge conservation we can expect that the
chiral field $\Phi^{k}$ would get renormalized according to
\begin{equation}
\tilde{\Phi}^{k}=\Phi^{k}+ A^{k-2}_{pk}g\Phi^{k-2}+
A^{k+2}_{pk}g^{*}\Phi^{k+2}+ A^{k}_{p\bar{p}k}\Phi^{k}gg^{*}+ O(g^{3})\,.
 \label{c1}
\end{equation}

where the barred indicies correspond to the antichiral coupling constants and
fields. The requirement that $\partial_{k}G_{ij}=0$ and the fact that the
three point function of chiral fields
 $\langle\Phi^{k}\Phi^{*k'}\Phi^{p}\rangle=0$ for any $k,k'$, sets the
constants $ A^{k-2}_{pk},A^{k+2}_{pk}$ to zero.

 We can restrict   our attention to the second order derivatives of metric
tensor which for the perturbed theory   (~\ref{13}) is given by
 \begin{eqnarray}
 \partial_{p}\partial_{\bar{p}}G_{k\bar{k}}= & 2A^{k}_{p\bar{p}k}& +
\label{c2} \\
 \frac{1}{(2\pi)^{2}} \int _{ |z_{1}-w_{1}|^{2}=1 }
& d^{2}Z_{2}d^{2}W_{2}^{*}&
\left |
\langle\Phi^{k}(Z_{1})\,\Phi^{*k}(W^{*}_{1})\,\Phi^{p}(Z_{2})\,\Phi^{*p}(W^{*}_{2})\rangle
\right |^{2}
\nonumber
\end {eqnarray}

where we set also $\theta _{z_{1}}\theta^{*}_{w_{1}}=0$, which allows us to
keep
the normalization of the Zamolodchikov metric \cite{rg:fl8} at the fixed point
independent of the dimension of the field $\Phi^{k}$.

For the Riemanian geometry the curvature tensor can be expressed  in this case
as
\begin{equation}
R_{abcd}=-\frac{1}{2}(G_{ac,bd}+G_{bd,ac}-G_{ad,bc}-G_{bc,ad})
\label{c3}
\end{equation}
For example   the curvature component $R_{k\bar{p}p\bar{k}}$ is given
by

\begin{equation}
R_{k\bar{p}p\bar{k}}=-\frac{1}{2}(G_{kp,\bar{k}\bar{p}}+G_{\bar{k}\bar{p},kp}
-G_{p\bar{p},k\bar{k}}-G_{k\bar{k},p\bar{p}}).
\label{c4}
\end{equation}

In the bare coordinate system  this would read

\begin{equation}
R_{k\bar{p}p\bar{k}}=\frac{1}{2}(I_{0}+\tilde{I}_{0} -I_{2})
\label{c5}
\end{equation}
where
\begin{eqnarray}
I_{0}&=& \nonumber \\
\frac{1}{(2\pi)^{2}} \int _{ |z_{1}-w_{1}|^{2}=1 }
& d^{2}Z_{2}d^{2}W_{2}^{*}&
\left |
\langle\Phi^{k}(Z_{1})\,\Phi^{*k}(W^{*}_{1})\,\Phi^{p}(Z_{2})\,\Phi^{*p}(W^{*}_{2})\rangle
\right |^{2} \nonumber \\
&=&I(k,p) \label{c6} \\
\tilde{I}_{0}&=&I(p,k) \label{c77}
\end{eqnarray}

\begin{eqnarray}
I_{2}&=& \nonumber \\
\frac{1}{(2\pi)^{2}} \int _{ |z_{1}-w_{1}|^{2}=1 }
& d^{2}Z_{2}d^{2}W_{2}^{*}&
\left |
\langle\Phi^{k}(Z_{1})\,\Phi^{*k}(Z^{*}_{2})\,\Phi^{p}(W_{1})\,\Phi^{*p}(W^{*}_{2})\rangle
\right |^{2} \nonumber \\
&=&0 \, . \label{c7}
\end{eqnarray}

The four point functions of equation  (~\ref{c6}) and their integrals  $I_{0}$
are calculated in the Appendix.

The fact that the integral $I_{2}=0$
follows from considering the four point function of $I_{2}$ with lowest
componets  of $\Phi^{p},\Phi^{k}$ fields. It is  equal to total
derivative in $w_{2}$ as can be easily checked using the supersymmetry (as
in the $I_{0}$ case) and its integral is   zero. In fact one could expect
such result  because  the two point functions of chiral-chiral
field vanish due to N=2 SUSY and so there could not be a sypersymmentric result
of integral $I_{2}$.

Let's  further consider what limits are imposed by the U(1) symmetry on the
curvature tensor  at $g=0$ point i.e. how can it be changed by the allowed
coordinate transformation. Because of the U(1) symmetry the commponents of
the curvature tensor can not be changed by a linear transformation of
coordinates as one can not freely rotate one coordinate into the other. Since
in our perturbation theory (~\ref{13}) we are limited only to the third order
in
$g$ coordinate transformation it means that the curvature tensor is restricted
to have the value it has in the bare coordinate system as calculated in
equation (~\ref{c5}). It means that if we allow the second order in $gg^{*}$
renormalization of chiral fields the curvature  tensor would have to be
\begin{equation}
R_{k\bar{p}p\bar{k}}=\frac{1}{2}(I_{0}+\tilde{I}_{0} +2A^{k}_{p\bar{p}k}+
2A^{p}_{k\bar{k}p})
\label{c9}
\end{equation}
where constants are related to the renormalization of the fields by the
relation
\begin{eqnarray}
\tilde{\Phi}^{p}&=&\Phi^{p} + A^{p}_{k\bar{k}p}g^{k}g^{*k}\Phi^{p} \label{c10}
\\ \tilde{\Phi}^{k}&=&\Phi^{k} + A^{k}_{p\bar{p}k}gg^{*}\Phi^{k} \label{c11}
\end{eqnarray}
This for  $k=p$   means that the second order renormalization of the
$\Phi^{p}$ field must be  zero (if there are no first order terms) because the
curvature is a tensor and must transform homogeneously. Therefore we need to
look more closely at what form of chiral fields renormalization is allowed.

It is possible to better understand  and confirm this result if we adopt the
method of looking at renormalization of the conformal fields as the
  coupling constant dependent contact  terms in the operators
product expansion of chiral fields  which was developped by Kutasov
\cite{rg:cr1}.

Since we are only interested in the four point function of
$\langle \Phi_{k} \Phi_{*k} \Phi_{p} \Phi_{*p}\rangle$ or in other words the
second order derivatives of $G_{k\bar{k},p\bar{p}}$ or $G_{p\bar{p},k\bar{k}}$
we can restrict our attention to the contact terms which arise from the
expansion of two chiral fields

\begin{eqnarray}
\Phi_{k}(Z_{1})\Phi_{p}(Z_{2})&=& (conformal\,\, theory)
+\delta^{2}(Z_{12})C^{k-2}_{kp}\Phi_{k-2}(Z_{2}) \nonumber \\
&+&\delta^{2}(Z_{12})C^{k}_{kp}g^{*}\Phi_{k}
(Z_{2})+\delta^{2}(Z_{12})C^{p}_{kp}g^{*k}\Phi_{p}(Z_{2})
\label{c12}
\end{eqnarray}

where $\delta^{2}(Z_{12})=\delta^{2}(\theta_{1}-\theta_{2})\delta^{2}(z_{12})$
is the SUSY delta function and the $C^{i}_{jl}$ constant can depend on
$gg^{*},\,\,g^{k}g^{*k}$. In principle we could have also the contact terms
from
the OPE of chiral-antichiral field but it is easy to check that  such   contact
terms could contribute to the second order derivatives of  $G_{ij}$ by  zeroth
order in $g$ components and these have to be zero.  We can read it from  the
expansion $\Phi^{k}(Z_{1})\Phi^{*p}(Z^{*}_{2})\approx \delta^{2}(z_{12})
C(g=0)\phi^{k-p}_{l}(z_{2},\theta_{1},\theta^{*}_{2})$ ( where $\phi^{k-p}_{l}$
is a nonchiral field). At $g=0$ conformally invariant point the left and right
hand side of this realtion should have the same dimensions and since the
$\frac{k+p}{2(p+2)} < 1$ for any $k$ there can not be such a term as always the
dimension of the right hand side would be greater than 1.

We can rexpress the second order derivatives of the metric tensor by the
contact terms as
\begin {eqnarray}
\partial_{p}\partial_{\bar{p}}\langle\Phi_{k}(Z)\Phi^{*}_{k}(0)\rangle&=&
I_{0}+|C^{k-2}_{kp}|^{2}\langle\Phi_{k-2}(Z)\Phi^{*}_{k-2}(0)\rangle \nonumber
\\ &+&
(\overline{C^{k}_{kp}}+C^{k}_{kp})\langle\Phi_{k}(Z)\Phi^{*}_{k}(0)\rangle
\label{c13} \end{eqnarray}
so that we can identify (at $g=0$)
the first order derivatives of the contact term
constants with the second order in $g$ renormalization constants for the
 fields
$(C^{k}_{kp}g^{*})_{,\bar{p}}=C^{k}_{kp}=A^{k}_{p\bar{p}k}$ and
 $C^{k-2}_{kp}=0$.

Following the method developped by   Kutasov \cite{rg:cr1} the first order
derivatives of the contact term of the expansion (~\ref{c12}) can be determined
by the four point function in the frame in which the first order derivatives of
the metric are set to zero. More precisely they are equal to the terms which
are
proportional to the delta type singularities arising from the once integrated
four point function \cite{rg:cr1}. In our case we have to consider the contact
term  coming from

\begin{eqnarray}
\frac{1}{(2\pi)} \int
& d^{2}W_{2}^{*}&
\left |
\langle\Phi^{k}(Z_{1})\,\Phi^{*k}(W^{*}_{1})\,\Phi^{p}(Z_{2})\,\Phi^{*p}(W^{*}_{2})\rangle
\right |^{2}= \nonumber \\
&=&(finite\,\,term)+\delta^{2}(Z_{12})C^{k}_{kp}
\langle\Phi_{k}(Z_{1})\Phi^{*}_{k}(W^{*}_{1})\rangle
\label{c14}
\end{eqnarray}

Let's consider the lowest components of $\Phi_{k}$ field four point function
which is calculated in the Appendix

\begin{eqnarray}
&\langle & \phi^{k}(z_{1})\,\phi^{*k}(w_{1})\,\psi^{p}(z_{2})\,
\psi^{*p}(w_{2})\rangle     = \nonumber \\
&=&\frac{4}{|z_{12}|^{2}} [ 2\epsilon
\langle \phi^{4}\rangle |^{2}+\partial_{\bar{w_{2}}}( 2\epsilon
\langle\phi^{4}\rangle)\overline{(w_{2}-z_{1})\langle\phi^{4}\rangle} +c.c
\nonumber \\
 &+& |\partial_{\bar{w_{2}}}(w_{2}-z_{1}) \langle \phi^{4}\rangle |^{2} ]
 + tot. derivatives
\label{c15}
\end{eqnarray}

where now the total derivatives terms include the dependence
$\partial_{w_{2}}\frac{1}{\bar{w}_{2}-\bar{z}_{i}}=i\pi
\delta^{2}(w_{2}-z_{i})$ \cite{rg:fl19}, and $\langle\phi^{4}\rangle=
\langle\phi^{k}(z_{1})\,\phi^{*k}(w_{1})\,\phi^{p}(z_{2})\,
\phi^{*p}(w_{2})\rangle$.
Integrating over the $w_{2}$   function (~\ref{c15})
we find that  the contact term dependence gives the zero coefficient in front
of the  $\delta^{2}(Z_{12})$ term. The contribution from integrating the
$\langle\phi^{4}\rangle$ term is also finite as in the limit
$z_{12}\rightarrow 0$. The last statement can be checked using the  known
formula  \cite{rg:fl22}, \cite{rg:fl25}

\begin{eqnarray}
&\int& d^{2}z| z|^{2a}|1-z|^{2b}|\zeta-z|^{2c}
 =\frac{s(a+b+c)s(b)}{s(a+c)} |J_{1}(a,b,c,\zeta)|^{2}+\nonumber \\
&+ & \frac{s(a)s(c)}{s(a+c)} |J_{2}(a,b,c,\zeta)|^{2}
\label{c16}
\end{eqnarray}

Here
\begin{eqnarray}
J_{1}(a,b,c,\zeta) &=&
\frac{\Gamma(-a-b-c-1)\Gamma(b+1)}{\Gamma(-a-c)}\times \nonumber \\
& \times & F(-c,-a-b-c-1,-a-c,\zeta) \nonumber \\
J_{2}(a,b,c,\zeta)&=&
z^{1+a+c}\frac{\Gamma(a+1)\Gamma(c+1)}{\Gamma(a+c+2)} F(-b,a+1,a+c+2,\zeta)
\nonumber \\
\,\,
\end{eqnarray}
  $F$ is the hypergeometric function, and $s(a)=sin(\pi a)$.

 We find that the integration  gives
(with $z_{1}=1$ and $w_{1}=0$ for clarity)

\begin{eqnarray}
&(4\epsilon)^{2}&|z_{2}-1|^{-2(1-k/(p+2))}|z|^{_2k/(p+2)}
 \frac{s(\frac{p}{p+2}) s(\frac{k}{p+2})}{s(\frac{p+k}{p+2})} \times
\nonumber\\
 &\times&  \left[ \left|
\frac{\Gamma(-1+\frac{p}{p+2})\Gamma(1+\frac{k}{p+2})}
{\Gamma(\frac{p+k}{p+2})}F(\frac{p}{p+2},-1+\frac{p}{p+2},\frac{p+k}{p+2},1-z)
\right|^{2}- \right. \nonumber \\
&\,&\left.\left| \frac{\Gamma(1-\frac{p}{p+2})\Gamma(1-\frac{k}{p+2})}
{\Gamma(2-\frac{p+k}{p+2})}z^{1-\frac{p+k}{p+2}}
F(-\frac{k}{p+2},1-\frac{k}{p+2},2-\frac{p+k}{p+2},1-z)
 \right|^{2}\right] \nonumber \\
\,\, .
\label{c17}
\end{eqnarray}
which behaves as $|z-1|^{-2(1-\frac{k}{p+2})}$ when $z \rightarrow 1$ and is no
singular enough to give contribution to delta type singularity (integral over
a small neighborhood  around 1 goes to zero).

{}From the contact terms  consideration we find the same result we concluded
before. The SUSY demands that the second order in $|g|$ renormalization of the
chiral fields is zero.

One more way of understanding this result is to consider how the metric tensor
components $G_{p\bar{p}}$ would have to change under the allowed coordinate
transformation $\tilde{g}=g(1+agg^{*})$. This kind of transformation would
have to produce the nonzero component $\tilde{G}_{pp}=-ag^{*}g^{*}$ wich is not
compatible with the chiral renormalization of the fields (~\ref{c1}). This
means
that such a coordinate transformation have to be excluded.

{}From the
Callan-Szymanzik equation (~\ref{19}) we find that  the anomalous
dimensions matrix and the beta function remain unchanged.
\begin{eqnarray}
\gamma ^{i}_{k}=\Delta_{k}
\delta^{i}_{k} \label{43}
\\ \beta (g)=\epsilon g
\label{44}
\end{eqnarray}

\section{ Discussion}

Our calculation shows that there is no perturbative fixed point near the
 $A_{p}$ fixed point theory \cite{rg:fl2}, \cite{rg:fl3}, \cite{rg:fl12},
\cite{rg:fl13}.  As we have mentioned in the introduction such result is well
understood on the consideration of the Witten index $Tr(-)^{F}=p$ of N=2 SUSY
theory, which can not be changed perturbatively. We know \cite{rg:fl2},
\cite{rg:fl12}, \cite{rg:fl13} that in the infrared limit the
theory given by the superpotential $W(\Phi)=\Phi^{p+2} + g\Phi^{p}$ will flow
to the fixed point theory $W(\Phi)=\Phi^{p}$ so these two fixed points should
be
infinitely apart as was argued by Cveti\v{c} and Kutasov \cite{rg:fl7}. Our
result for the $\beta $ function $\beta= g\epsilon $
confirms that to the leading order in $g$. It states
that the nonrenormalization theorem holds in this perturbation theory in a
simple way $\tilde{\Phi}^{k}=\Phi^{k}$. The most important content of it is
that the chiral renormalization of the fields is not compatible with the
allowed (by U(1) charge conservation) redefinition of the coupling constants
$\tilde{g}=g(1+agg^{*})$.

This information can be also read of from the expansion of the Zamoldchikov
metric $G_{ij}$ in the bare coordinates
$G^{0}_{k\bar{k}}=\delta_{k\bar{k}}(1+I(k,p))$ where $I(k,p)=1+O(\epsilon)$.
There are no singular terms in the expansion of the metric as $\epsilon
\rightarrow 0$. The chiral bare couplings coordinate system gives a well
defined metric  along the renormalization group flow  given by the
perturbed potential $W(\Phi)=\Phi^{p+2} + g\Phi^{p}$. There is no need for the
singular coupling redefinitions to keep the metric regular in the
neighborhood of the fixed point as it is the case in N=0, N=1 SUSY L-G models
\cite{rg:fl15}, \cite{rg:fl3}.

Our result is also in agreement with the recent work by West and Howe
\cite{rg:cr2}. It was shown there that as long as the SUSY is preserved by the
perturbation the effective potential does not get renormalized. However there
is the possibility that the nonrenormalization theorem would not hold in the
more general renormalization prescription scheme \cite{rg:cr2}. The present
work states the nonrenormalization theorem even more restrictively,
showing that there are no compatible with N=2 SUSY renormalization constant
terms of order $gg^{*}$ other than zero.

We feel also obliged to comment on the result obtained by Leaf-Hermann
\cite{rg:fl25} for the integration of the four point function
$\langle \psi^{p}\psi^{*p}\psi^{p}\psi^{*p} \rangle$ over two variables.
Clearly his integral is the special case of our integration formulas  when
$k=p$ $(\epsilon_{k}=\epsilon)$. The discrepancy between our
$I_{h\bar{h}}\approx4\pi^{2}$ and his result $I_{h\bar{h}}(L-H)\approx4\pi^{2}
\epsilon$ is quite substantial and so deserves a full comment.
 The four point functions we integrate are identical (however obtained by
different methods) as can be easily checked, so the source of the difference
lies in the method of integration.
 The problem is caused by the expansion of the
$_{2}F_{1}(2\epsilon,2\epsilon;4\epsilon;z)$ and
$_{2}F_{1}(1-2\epsilon,1-2\epsilon;1-4\epsilon;z)$ hypergeometric functions of
 \cite{rg:fl25}. In general any function with the parameters
$_{2}F_{1}(\alpha,\alpha +m;\gamma;z)$  where $m$ is natural number, will have
a
very nonuniform expansion \cite{rg:fl26}, and additionally when $2\alpha
+m=\gamma$ there will be also a  logarithmic  divergence at $z=1$
\cite{rg:fl23}. Consequently for the above set of parameters the hypergeometric
function contains the infinite series of logarithms \cite{rg:fl26}, and the
approximation of $_{2}F_{1}$ functions for small $\epsilon$, which we believe
was made in the \cite{rg:fl25} is not valid. Neglecting the fact of highly
ununiform
behaviour of this type of hypergeometric functions would probably always lead
to the wrong answer. Analytic continuation
 of the complete 2D integrals will lead to the correct answer free of
divergences for every term in $\epsilon$ expansion as we have shown in the
appendix.

I would like to thank Emil Martinec for bringing this problem to my attention
and spending considerable amount of time on discussions. I would like also to
aknowledge the referee for constructive comments (specially pointing out the
nonuniform behaviour of some hypergeometric functions) which led us to
 correct some major flaws in an early version of this paper.
This paper is submitted in partial fulfillment of the requirements for
 a Ph.D. degree in physics at the University
 of Chicago.

 \section{Appendix}

 The four point function for the highest components
of  $\langle \Phi^{k}\,\Phi^{*k}\,\Phi^{p}\,\Phi^{*p}\rangle$ can be easily
calculated in terms of the lowest components using the OPE with the
supercurrents $G^{\pm}_{\frac{1}{2}}$ \cite{rg:fl6}, \cite{rg:fl18}. For
clarity
in the formulas of this section we introduce the decomposition of (anti)chiral
fields in terms of the lowest $\varphi^{k}$ and highest $\psi^{k}$ components,
 \begin {eqnarray}
\Phi^{k}(Z) &=& \varphi^{k}(z)+\theta\psi^{k}(z) \nonumber \\
\Phi^{*k}(Z^{*}) &=& \varphi^{*k}(z)+\theta^{*}\psi^{*k}(z)
\label{36}
\end{eqnarray}
and mostly suppress antiholomorphic dependence.

The relative normalization of $\varphi^{k}$ and $\psi^{k}$ is given by the
two point function (~\ref{12}). Then the four point function of highest
components is equal to
\begin{eqnarray}
& &
\langle\psi^{k}(z_{1})\,\psi^{*k}(w_{1})\,\psi^{p}(z_{2})\,\psi^{*p}(w_{2})\rangle=
\nonumber \\
& = & \frac{4}{(z_{1}-z_{2})}[-\frac{p}{p+2} \partial_{w_{1}} +
\frac{k}{p+2} \partial_{w_{2}}
+(w_{1}-w_{2})\partial_{w_{1}}\partial_{w_{2}}]\times \nonumber \\
& \times &
\langle \varphi^{k}(z_{1})\,\varphi^{*k}(w_{1})\,\varphi^{p}(z_{2})\,
\varphi^{*p}(w_{2})\rangle
\label{37}
 \end{eqnarray}
Using the fact that
$\varphi^{p}= :exp\left( ip\varphi / \sqrt{2p(p+2)} \right) :$
one can easily find the formula for the four point function up to  contact
terms
\begin{eqnarray}
& &
\langle\psi^{k}(z_{1})\,\psi^{*k}(w_{1})\,\psi^{p}(z_{2})\,\psi^{*p}(w_{2})\rangle=
\nonumber \\
&=& \frac{4k}{p+2}\frac{(z_{1}-w_{1})^{-\frac{k}{p+2}}
(z_{2}-w_{2})^{-\frac{p}{p+2}} }{(z_{1}-z_{2})(w_{1}-w_{2})}
\left[
\frac{(z_{1}-z_{2})(w_{1}-w_{2})}{(z_{1}-w_{2})(w_{1}-z_{2})} \right]
^{\frac{k}{p+2}} \nonumber \\
&\times &
\left[1+\frac{p}{p+2}
\frac{(z_{1}-w_{2})(w_{1}-z_{2})}{(w_{2}-z_{2})(w_{1}-z_{1})} + \right .
\nonumber \\
&+& \left .
\frac{k}{p+2}[-1
+\frac{(w_{2}-z_{2})(w_{1}-z_{1})}{(z_{1}-w_{2})(w_{1}-z_{2})}] \right]
\label{38}
\end{eqnarray}
Analogously we can calculate the contribiution for the sypersymmetric four
point function from the lowest components of chiral fields.

\begin{eqnarray}
& &
\langle\phi^{k}(z_{1})\,\phi^{*k}(w_{1})\,\psi^{p}(z_{2})\,\psi^{*p}(w_{2})\rangle=
\nonumber \\
& = & \frac{2}{(z_{1}-z_{2})}[-\frac{p}{p+2}  +
 (z_{1}-w_{2})\partial_{w_{2}}
]\times \nonumber \\
& \times &
\langle \varphi^{k}(z_{1})\,\varphi^{*k}(w_{1})\,\varphi^{p}(z_{2})\,
\varphi^{*p}(w_{2})\rangle = \nonumber \\
&=& 2\frac{(z_{1}-w_{1})^{-\frac{k}{p+2}}
(z_{2}-w_{2})^{-\frac{p}{p+2}} }{(z_{1}-z_{2})}
\left[
\frac{(z_{1}-z_{2})(w_{1}-w_{2})}{(z_{1}-w_{2})(w_{1}-z_{2})} \right]
^{\frac{k}{p+2}} \nonumber \\ &\times &
\left[-\frac{p}{p+2}
\frac{(z_{1}-z_{2})}{(z_{2}-w_{2})}+
\frac{k}{p+2}\frac{(w_{1}-z_{1})}{(w_{2}-w_{1})}
\right]
\label{39}
\end{eqnarray}

The integral of (~\ref{38}) or (~\ref{39})  over  $Z_{2}$ and $W_{2}$
has various poles of order close to  2 and logarithmic ones. These poles
are the consequence of the OPE of any two chiral operators which produces the
pole under the integration whenever two (or more) operators come close
together. Logarithmic poles are the consequence of the presence of the
U(1) current in the OPE of chiral-antichiral fields.  Due to
SUSY there are contact terms \cite{rg:fl19} which exactly cancel these  poles.
This comes from a more careful consideration of the
OPE of the holomorphic and antiholomorphic part of  chiral fields.

 The OPE of the holomorphic part of the chiral field, given in terms of the
lowest and higest components, normalized by the two point function (~\ref{12}),
is
 \begin{eqnarray}
\varphi^{k}(z_{1})\varphi^{*k}(z_{2})&=& \frac{1}{ z_{12}^{2\Delta}}
[1 +z_{12}\frac{2q}{\frac{2}{3}c}J(z_{2}) + \cdots] \nonumber \\
\psi^{k}(z_{1})\psi^{*k}(z_{2})&=& \frac{4\Delta}{z_{12}^{2\Delta +1}}
[1 +z_{12}\frac{2q-1}{\frac{2}{3}c}J(z_{2}) + \cdots]
\label{40}
\end{eqnarray}
The antiholomorphic parts have the same OPE. In expansion ~\ref{40} we write
down
only terms which  produce poles under the integral. A
logarithmic pole arises from the OPE of the charge current $J(z)$ with a third
 operator present in the four point function . $J\bar{J}$ is the lowest
component of a superfield;
therefore a supersymmetric renormalization procedure should cancel these
divergences.
Analytic continuation of the hypergeometric function which arises from
integration of (~\ref{38}) automaticaly subtracts these divergences and is
compatible with N=2 supersymmetry, as our calculation shows.

 To evaluate the integral (~\ref{c6}) we calculate the contribution given by
diffrent components of superchiral field $\Phi^{k}$. All the formulas
involved can be written  more clearly if we use  the translational invariance
and choose the scale by putting down $z_{1}=1\,\,, w_{1}=0$. It is also
convenient to introduce new coordinates
\begin{equation}
\zeta=\frac{z_{2}(w_{2}-1)}{w_{2}(z_{2}-1)}\, , \,\,\,\,
\eta=\frac{w_{2}-1}{w_{2}} .
\label{48}
\end{equation}
With these simplifications the
contributing parts of integral (~\ref{c6}) look like

\begin{eqnarray}
I_{h\bar{h}} &=& \int
d^{2}\zeta d^{2}\eta  \left | 4 \frac{k}{p+2}
(\zeta-\eta)^{-2\epsilon}(1-\zeta)^{-1+2\epsilon}\zeta ^{-1+2\epsilon_{k}}
(1-\eta)^{-2\epsilon} \eta^{-1+2\epsilon} \right | ^{2}  \nonumber \\
 & \times & \left | 1 +\frac{p}{p+2}\frac{\zeta}{(1-\zeta)}+
\frac{k}{p+2}[-1 + \frac{(1-\zeta)}{\zeta}] \right |
^{2}
\label{49} \\
I_{h\bar{l}} &=& \int
d^{2}\zeta d^{2}\eta    \frac{2k}{p+2}\left | 2
(\zeta-\eta)^{-2\epsilon}(1-\zeta)^{-1+2\epsilon}\zeta ^{-1+2\epsilon_{k}}
(1-\eta)^{-2\epsilon} \eta^{-1+2\epsilon} \right | ^{2}  \nonumber \\
 & \times & \left ( 1 +\frac{p}{p+2}\frac{\zeta}{(1-\zeta)}+
\frac{k}{p+2}[-1 + \frac{(1-\zeta)}{\zeta}] \right ) \times \nonumber \\
& \times &
 \left ( \frac{p}{p+2}\frac{1}{(1-\bar{\zeta)}}-
\frac{k}{p+2} \right )
\label{50}\\
I_{l\bar{l}} &=& \int
d^{2}\zeta d^{2}\eta   \left | 2
(\zeta-\eta)^{-2\epsilon}(1-\zeta)^{-1+2\epsilon}\zeta ^{-1+2\epsilon_{k}}
(1-\eta)^{-2\epsilon} \eta^{-1+2\epsilon} \right | ^{2} \times  \nonumber \\
& \times & \left | \frac{p}{p+2}\frac{1}{(1-\zeta)}-
\frac{k}{p+2} \right |
^{2}
\label{51}
\end{eqnarray}
The indexes $l,(\bar{l})$ and $h,(\bar{h})$ refer respectively to the lowest
and highest left (right) components of the chiral superfields $\Phi^{k}$,
$\Phi^{*k}$.

To evaluate these integrals we follow the method developed by Kawai, Lewellen
and Tye \cite{rg:fl20} to calculate  closed string amplitudes. Here we can
almost directly use the formula for the five point function which has exactly
the same form as the above integrals. After a simple
change of integration variables we can express the KLT \cite{rg:fl20} formula
(3.26) in
 a simple way using the generalized hypergeometric functions. The  most
general form of this expression is

 \begin{eqnarray}
J &=& \int d^{2}z_{1}d^{2}z_{2}|
z_{1}|^{2a_{1}}|1-z_{1}|^{2b_{1}}|z_{1}-z_{2}|^{2c}
|z_{2}|^{2a_{2}}|1-z_{2}|^{2b_{2}} \times \nonumber \\
& \times &
z_{1}^{m_{1}}(1-z_{1})^{n_{1}}z_{2}^{m_{2}}(1-z_{2})^{n_{2}} \nonumber \\
&=&s  (a_{1})s (b_{2})
\int_{0}^{1} d \zeta _{2}\int_{0}^{\zeta _{2}} d\zeta_{1}
|\zeta_{1}|^{\tilde{a_{1}}}|1-\zeta_{1}|^{\tilde{b_{1}}}
|\zeta_{1}-\zeta_{2}|^{c}
|\zeta_{2}|^{\tilde{a_{2}}}|1-\zeta_{2}|^{\tilde{b_{2}}}  \nonumber \\
& \times & \int_{-\infty}^{0} d \eta _{1} \int_{1}^{\infty} d \eta_{2}
|\eta_{1}|^{a_{1}}|1-\eta_{1}|^{b_{1}}|\eta_{1}-\eta_{2}|^{c}
|\eta_{2}|^{a_{2}}|1-\eta_{2}|^{b_{2}} \nonumber \\
&+& (1 \longleftrightarrow 2)
 \label{52}
\end{eqnarray}

Where
\begin{eqnarray}
m_{i},\,\,n_{i} \,\, \in  Z \,\, i=1,2 \nonumber \\
\tilde{a_{i}}=a_{i}+m_{i},\,\,\,\tilde{b_{i}}=b_{i}+n_{i},
\label{53}
\end{eqnarray}
 $ (1 \longleftrightarrow 2)$ means exchange of indexes 1, 2 and
 $s(a)=sin(\pi a)$.

After a change of integration variables
\begin{equation}
\eta_{1} \rightarrow \frac{1}{1-\eta_{1}}
\label{54}
\end{equation}
we can express the double integral in $\zeta_{i}$ and $\eta_{i}$ as
$_{3}F_{2}$ generalized hypergeometric functions \cite{rg:fl21},
\cite{rg:fl22}.
 Using the integral
definitions of hypergeometric functions \cite{rg:fl21},
\cite{rg:fl22} and standard transformation
formulas of them  \cite{rg:fl21},
\cite{rg:fl22} we find the result for $J$ to be
\begin{eqnarray}
J &=&
\pi^{2}\frac{\Gamma(\tilde{a_{1}}+1)\Gamma(\tilde{b_{2}}+1)
\Gamma(\tilde{a_{2}}+\tilde{a_{1}}+c+2)\Gamma(c+1)}
{\Gamma(\tilde{a_{1}}+c+2)
\Gamma(\tilde{b_{2}}+\tilde{a_{2}}+\tilde{a_{1}}+c+3)} \times \nonumber \\
& \times & _{3}F_{2}
(-\tilde{b_{1}},\tilde{a_{1}}+1,\tilde{a_{2}}+\tilde{a_{1}}+c+2;
\tilde{a_{1}}+c+2,\tilde{b_{2}}+\tilde{a_{2}}+\tilde{a_{1}}+c+3;1) \times
\nonumber \\
&\times& \frac{\Gamma(-1-b_{1}-a_{1}-c)\Gamma(-1-b_{2}-a_{2}-c)}
{\Gamma(-b_{1}-c)\Gamma(-a_{2}-c)\Gamma(-b_{2})\Gamma(-a_{1})} \times
\nonumber \\
&\times& _{3}F_{2}(-c,1+a_{1},1+b_{2};-c-a_{2},-c-b_{1};1) \nonumber \\
&+& (1 \longleftrightarrow 2)
 \label{55}
\end{eqnarray}
After plugging in the values of the exponents from (~\ref{49}), (~\ref{50}),
(~\ref{51}) into the equation (~\ref{52}) we get a not  very appealing
expresion
containing values of $_{3}F_{2}$ type hypergeometric functions \cite{rg:fl23}
at one (which we will call further $_{3}F_{2}$ series). The general summation
formula for $_{3}F_{2}$ series anologous to the Gauss summation formula for
"regular" hypergeometric functions,
$_{2}F_{1}(a,b;c;1)=\frac{\Gamma(c)\Gamma(c-a-b)}{\Gamma(c-a)\Gamma(c-b)}$,
is not known. There exist several special summation formulas \cite{rg:fl23} for
$_{3}F_{2}$ series  but none of the functions we get here does have a
closed form. So we can not give an exact result for our integrals. One could
expect that result we have so far is quite useless and no better
 that no answer at all. But luckily enough there are general transformation
formulas for $_{3}F_{2}$ series \cite{rg:fl23} which can be used here with
considerable
success.  Using the transformation formulas (4.3.1, 4.3.4, 4.3.4.2) given  in
Slater's book [23] we are able to transform  the  $_{3}F_{2}$ series involved
in
our calculation  into one another  and obtain much simplified  answers for the
integrals (~\ref{49}), (~\ref{50}), (~\ref{51}) :

\begin{eqnarray}
I_{l,\bar{l}} &=&I(k,p)= \Gamma^{2}(1+2\epsilon)\Gamma^{2}(1-2\epsilon)
\times \nonumber \\
&\times& [_{3}F_{2}^{2}(2\epsilon,-2\epsilon,2\epsilon_{k};1,1;1) +
\nonumber \\
&+& (2\epsilon)^{2}(1-2\epsilon_{k})^{2} \,
_{3}F_{2}^{2}(1+2\epsilon,1-2\epsilon,2-2\epsilon_{k};2,2;1) ]
\label{56} \\
I_{h,\bar{l}}&=& I_{l,\bar{h}}= I(k,p) \times 4(\Delta_{k}-\epsilon)
\label{57}\\
I_{h,\bar{h}} &=& I(k,p) \times \left [ 4( \Delta_{k}-\epsilon) \right ] ^{2}
\label{58}
\end{eqnarray}

 The above expressions are finite since  a generalized hypegeometric
function      $_{3}F_{2}(d_{1},d_{2},d_{3};e_{1},e_{2};z)$ is absolutely
convergent at $z=1$ if $(e_{1}+e_{2}-d_{1}-d_{2}-d_{3})>0$ \cite{rg:fl23},
which is
clearly   fullfiled for the $_{3}F_{2}$ series appearing here.
 For the $\epsilon \ll 1$ we have that $I(k,p)=1+O(\epsilon^{2})$ because
the $_{3}F_{2}(2\epsilon,-2\epsilon,2\epsilon_{k};1,1;1)$ is finite for any
$\epsilon$ and is 1 at $\epsilon=0$ and also
$_{3}F_{2}(1+2\epsilon,1-2\epsilon,2-2\epsilon_{k};2,2;1)$ is finite for all
$\epsilon$ as well as for $\epsilon=0$. The last statement is not abvious
but can be easly checked using the formula given by Luke  \cite{rg:fl24}

\begin{equation}
_{3}F_{2}(d,1,1;2,e;1)=\frac{e-1}{d-1}(\Psi(e-1)-\Psi(e-d))
\label{59}
\end{equation}
where $ \Psi (x)=\frac{d}{dx}  ln \Gamma(x)$.

 Unfortunately we
were unable to find generally valid transfomation of the expression (~\ref{55}
) which would lead us more directly to the above answers.
However it is possible to do a couple of checks of the results we obtained.

First it can be easily seen that the four point functions (~\ref{38}) and
(~\ref{39}) contain a total divergence  with respect to $w_{2}$ pieces. So
following   common  sense  the integration of the four point function with and
without these divergencies should be the  same. This check is nontrivial
because the total divergence terms are in fact divergent as $w_{2} \rightarrow
\infty$ like $w_{2}^{2\epsilon}$ and if true  confirms that the analytic
continuation really holds.

For the lowest and highest components of chiral
fields we have the following four point functions, defined up to total
divergence term

\begin{eqnarray}
& &
\langle\phi^{k}(z_{1})\,\phi^{*k}(w_{1})\,\psi^{p}(z_{2})\,\psi^{*p}(w_{2})\rangle=
\nonumber \\
&=& 2\epsilon \frac{2(z_{1}-w_{1})^{-\frac{k}{p+2}}
(z_{2}-w_{2})^{-\frac{p}{p+2}} }{(z_{1}-z_{2})}
\left[
\frac{(z_{1}-z_{2})(w_{1}-w_{2})}{(z_{1}-w_{2})(w_{1}-z_{2})} \right]
^{\frac{k}{p+2}} + \nonumber \\
&+& \partial _{w_{2}}(\ldots)
 \label{60}
\end{eqnarray}

 \begin {eqnarray}
& &
\langle\psi^{k}(z_{1})\,\psi^{*k}(w_{1})\,\psi^{p}(z_{2})\,\psi^{*p}(w_{2})\rangle=
\nonumber \\ &=&
2\epsilon\frac{4k}{p+2}\frac{(z_{1}-w_{1})^{-1-\frac{k}{p+2}}
(z_{2}-w_{2})^{-\frac{p}{p+2}} }{(z_{1}-z_{2})} \left[
\frac{(z_{1}-z_{2})(w_{1}-w_{2})}{(z_{1}-w_{2})(w_{1}-z_{2})} \right]
^{\frac{k}{p+2}} \nonumber \\  &\times &
\left[1
-\frac{(w_{2}-z_{2})(w_{1}-z_{1})}{(w_{1}-z_{2})(w_{1}-w_{2})} \right] +
\partial_{w_{2}}(\ldots)
\label{61}
\end{eqnarray}

As a check  for our results we did calculate the four point function, with
total divergence term included and without it. In either case we obtain the
same result (~\ref{56}), (~\ref{57}) , (~\ref{58}). This way we also prove that
total divergences do not have to be included in analytic continuation procedure
even if they are possibly contributing a divergent term.

Another prove that our calculation is correct is the fact that we obtain a
supersymmetric answer. The inegrated supersymmetric four point
function has a compact form form
\begin{eqnarray}
& \int & d^{2}Z_{2}d^{2}W_{2}
\langle \Phi^{k}(Z_{1},\bar{Z}_{1})\,\Phi^{*k}(W_{1}\,\bar{W}_{1})\,
\Phi^{p}(Z_{2},\bar{Z}_{2})\,\Phi^{*p}(W_{2},\bar{W}_{2})\rangle =
\nonumber \\
& = &4\pi^{2} |z_{1}-w_{1}|^{-4(\Delta_{k}-\epsilon)} \left | 1 +
4(\Delta_{k} -
\epsilon)\frac{\theta_{z_{1}}\theta_{w_{1}}^{*}}{(z_{1}-w_{1})} \right | ^{2}
\times I(k,p)
\label{62}
\end{eqnarray}
 which is exactly a N=2 SUSY two point function of a chiral field with a
dimension and  charge equal to $(\Delta_{k}-\epsilon)$. After the above
checks we can belive that the result we got is correct.

 \end{document}